# Transcutaneous Interference Spinal Cord Stimulation: Leadfield-Based Pareto Optimization of Electrode Montages for Improved Focality

Mariko Teragiwa[1], Leonel E. Medina[2], Alonso Carvajal[2], Kanata Yatsuda[1], Wenwei Yu[1,3], Jose Gomez-Tames[1,3]*,

[1]Department of Medical Engineering, Graduate School of Engineering, Chiba University, Chiba 263-8522, Japan

[2]Facultad de Ingeniería, Universidad de Santiago de Chile, Avda. Víctor Jara 3659, Santiago, Chile

[3]Center for Frontier Medical Engineering, Chiba University, Chiba 263-8522, Japan

* Corresponding author:
Jose Gomez-Tames
jgomez@chiba-u.jp

**Abstract**
**Purpose:** This study investigates the feasibility of transcutaneous interferential spinal cord stimulation (tISCS), a novel non-invasive neuromodulation method, using temporal interference to enhance focality and comfort in spinal cord stimulation. The central research question is whether tISCS can achieve targeted activation of spinal cord circuits while reducing unwanted stimulation of skin and muscle tissues, which are common limitations of conventional transcutaneous spinal cord stimulation (tSCS).
**Methods:** A finite element model of the lower thorax was developed to simulate electric field distributions for various skin electrode montages. To address the computational bottleneck associated with high-resolution modeling and montage optimization, we implemented a leadfield-based Pareto optimization strategy to identify the electrode configuration that maximizes the electric field in the spinal cord and minimizes it in off-target tissues. tISCS montages were compared with tSCS montages in terms of focality and stimulation efficiency.
**Results**: Optimized tISCS configurations significantly reduced electric field intensity in the skin by over 20-fold compared to tSCS. The ratio of spinal cord to skin electric fields increased by at least 10-fold, indicating enhanced focality. The injection current efficiency in tISCS can be leveraged to increase spinal cord electric fields by at least 5-fold while keeping skin exposure below the levels observed with tSCS.
**Conclusion:** tISCS enables deeper and more selective spinal cord stimulation compared to tSCS, with substantially reduced off-target effects. This is the first computational demonstration of tISCS feasibility. Leadfield-guided Pareto optimization enables efficient montage selection, providing a foundation for future experimental applications.

# 1. Introduction

One method for spinal cord stimulation involves implanting electrodes in the epidural space that deliver controlled electrical pulses to targeted areas of the spinal cord, known as epidural spinal cord stimulation (eSCS). eSCS is used to treat chronic and neuropathic pain and is being investigated to restore mobility for individuals with cervical spinal cord injuries [1], [2]. Despite its significant benefits, one challenge with eSCS is the inherent risk associated with implanting the stimulation leads and the pulse generator [3], [4], [5], [6]. Moreover, the placement of the stimulation leads may shift over time, potentially leading to unintended stimulation and diminishing the therapy's effectiveness. Another concern is that the leads could break or bend, resulting in a malfunction [7].

An alternative strategy is transcutaneous spinal cord stimulation (tSCS), which delivers electric current through surface electrodes placed on the skin over the spine, typically in the thoracolumbar region, generating an electric field intended to modulate spinal circuits [8]. However, tSCS is limited in its ability to precisely target specific neural structures due to substantial current attenuation at depth, resulting in low focality. Achieving sufficient intensity at the spinal cord requires high surface currents, which can activate cutaneous nociceptors-thereby causing pain-and unintentionally stimulate nearby organs, including muscles [8]. Therefore, minimizing activation or modulation of cutaneous nociceptors, sensory fibers, and/or motor fibers while maintaining effective spinal stimulation is a critical challenge.

One transcutaneous stimulation modality is the so-called temporal interference stimulation, which uses two pairs of electrodes to deliver kilohertz-frequency currents, producing a focused interferential area deep within tissues where neuromodulation would occur [9]. The targeted area is defined by a strong interference pattern resulting from the added effect of both current pathways, with modulation of nerve fiber activity purportedly occurring at the frequency of the envelope of the interference signal. Conversely, superficial areas are apparently predominantly influenced by the high-frequency components of the stimulation signals, which would result in no activation effects and/or conduction block on cutaneous afferents [10]. In the brain, temporal interference has been shown to modulate deep brain regions and is well tolerated [11][12]. For peripheral nerves, temporal electrical stimulation allows higher current delivery without exceeding pain thresholds, enabling deeper stimulation with improved comfort [13], [14].

However, it remains unclear whether temporal interference stimulation can effectively target the spinal cord with sufficient focality while minimizing cutaneous pain and muscle activation. This uncertainty presents a significant obstacle to translating temporal interference into SCS applications. A promising strategy for addressing whether temporal interference stimulation can effectively target the spinal cord is the utilization of computational modeling of electric fields in anatomically based models. Such models have proven valuable for optimizing montages in tSCS [15][16][17], as well as for investigating temporal interference in the brain [18]. Yet, to date, no studies have applied computational modelling specifically to transcutaneous interference spinal cord stimulation (tISCS). A major limitation in applying these models to tISCS is the substantial computational cost associated with high-resolution spinal cord models (e.g., 2.5 hours per montage [19]), which increases dramatically when optimizing. To overcome this bottleneck, the leadfield matrix approach offers an effective solution in neurostimulation [20], [21]. This approach involves precomputing and storing the electric field distributions generated by a set of electrodes pairs. Once this matrix is established, the resulting electric field for any arbitrary combination of the electrodes in the set can be rapidly obtained via linear combinations of the precomputed fields, enabling fast and efficient simulation of numerous stimulation configurations in temporal interference for the brain [22], [23]. Extending these modelling frameworks to tISCS would offer a critical foundation for exploring field distributions, targeting specificity, and enabling informed design of stimulation protocols before clinical implementation.

This study aims to computationally explore the feasibility of tISCS, focusing on its potential to reduce unwanted effects on the skin and muscle and to provide a more effective targeted approach. To this end, we estimated the electric field using an anatomically based finite element model of the thorax and optimized the electrode montage based on the Pareto front.

## 2. Methodology

### 2.1. Human model design

A realistic finite element method (FEM) model of the thorax was constructed based on. anatomical data from the lower thoracic spinal level (T11) [24] and surrounding anatomical tissues [25], [26]. The model was segmented into 12 tissues, including skin, fat, muscle, general thorax, bone, vertebrae, intervertebral disks, extradural tissue, dura matter, cerebrospinal fluid (CSF), spinal white matter (spinal-WM), and spinal gray matter (spinal-GM), as shown in the Figs. 1A and 1B. The electrical conductivity values were set according to Table 1 [27],[28],[29],[30],[31]. A conductivity tensor was implemented to account for the anisotropic nature of spinal-WM, where conductivity is higher along the caudal-to-rostral axis. tISCS operates at carrier frequencies of a few kilohertz, whereas tSCS typically uses frequencies below 100 Hz. While tissue conductivity is known to be frequency-dependent, we adopted the conductivity values commonly reported for tSCS. This approximation is justified, as prior study has shown that variations in conductivity within this frequency range have a negligible impact on tISCS [32].

## 2.2. Electrodes model

A two-layered electrode was used. The layer on top is silicone rubber (2401 mm², 44 S/m [33]). The bottom layer consists of a saline gel (2500 mm², 4 S/m [33]). It also includes a rectangular connector (108 mm², 1000 S/m) affixed underneath the rubber pad (Fig. 1C).

## 2.3. Electric field model

Using the quasi-static approximation and the finite element technique (COMSOL Multiphysics Ver. 6.2) to solve Laplace's equation [34], the electric potential produced by the current injected from a pair of electrodes (common bipolar tSCS) placed on the thorax was estimated using the following equation:

$$\nabla \cdot (\sigma \nabla \varphi) = 0, \qquad (2)$$

where $\varphi$ and $\sigma$ denote the scalar potential and tissue conductivity, respectively. The electric field $\vec{E} = -\nabla \varphi$ was calculated in each element. The electrode connections were treated as isopotential surfaces when applying the boundary conditions [35]. There were 7,139,950 tetrahedral elements. The top 0.01% outliers were removed from the sorted E-field intensity produced by potential mesh errors [36]. The results were numerically stable relative to a finer mesh.

Since the direction of the target nerve fibres, i.e., the dorsal column fibres, is assumed to align with the z-axis, only the z-directional component of the electric field within the spinal cord, which aligns with the driving force of fibre activation, was considered for the following analysis [37]. For bipolar tSCS, the electric field was obtained for a pair of electrodes. In the case of tISCS (Fig. 1D), the amplitude-modulated envelope ($E_{AM}$) is generated from the vector sum of the two resultant electric fields ($E_1$ and $E_2$), each corresponding to a pair of electrodes driven by an independent current source. The electric fields at location $\vec{r}$ are represented by $\vec{E_1}(\vec{r})$ and $\vec{E_2}(\vec{r})$. If $|\vec{E_1}| > |\vec{E_2}|$ and the angle $\alpha$ (angle between $|\vec{E_1}|$ and $|\vec{E_2}|$) is lower than 90°, the modulation depth is expressed by [18], [38]

$$|\vec{E}_{AM}(\vec{r})| = \begin{cases} 2|\vec{E_2}(\vec{r})| & \text{if } |\vec{E_2}(\vec{r})| < |\vec{E_1}(\vec{r})|\cos\alpha \\ 2\left|\vec{E_2}(\vec{r}) \times \frac{(\vec{E_1}(\vec{r}) - \vec{E_2}(\vec{r}))}{|\vec{E_1}(\vec{r}) - \vec{E_2}(\vec{r})|}\right| & \text{otherwise} \end{cases} \qquad (3)$$

The direction of one vector must be flipped if the angle is greater than 90°.

For the conventional tSCS, we used a bipolar montage with a midline orientation and inter-electrode distance of 2.5 cm [33][39]. The electrode positions for tISCS were determined through a quasi-optimization process, described in the next subsection. The total injected current was fixed at 2.5 mA for conventional tSCS and tISCS [40]. As a result, the current per electrode pair in tISCS is halved relative to conventional tSCS. To ensure a fair comparison, we also implemented a two-pair in-phase bipolar tSCS configuration (referred to as 2-tSCS), matching both the number of electrodes and the current density used in tISCS. The electrode positions for 2-tSCS were set identically to those used for tISCS. It is important to note that the choice of maximum injected current does not affect the generalizability of the results, as the electric field magnitude in the model can be scaled linearly with the injection current intensity.

2.4. Optimization method based on leadfield matrix

The leadfield matrix $\boldsymbol{A_n}$ is a linear mapping between the current injection $\boldsymbol{x_n}$ at electrode $n$ (relative to a reference) and the resulting electric field $\vec{E}$ $(p)$ at mesh node $p$, expressed as

$$\vec{E}\ (p) = \boldsymbol{A_n x_n}. \tag{4}$$

A total of $n$ = 162 electrode positions (6 vertical positions along z-axis × 27 horizontal positions around the thorax surfaces) were considered (Fig. 2A). The reference electrode was placed at the center of the thorax. . Thus, a total of 161 simulations were conducted to generate the matrix $\boldsymbol{A_n}$ using COMSOL Multiphysics (Ver. 6.2) on an Intel(R) Xeon(R) W-2295 processor clocked at 3.00 GHz and 128 GB of RAM. The computation took approximately 30 hours. Once $\boldsymbol{A_n}$ (161 montages × 7,139,750 no. model elements) was generated, the electric field for any linear combination of electrode currents could be predicted. Selecting a specific column from the leadfield matrix corresponds to selecting a specific electrode position. The color of each electrode in Fig. 2A matches the color of the corresponding column in the leadfield matrix. We chose to start optimization over a broad electrode placement area, covering the entire thorax, to identify a wide range of electrode configurations capable of producing strong interferential currents at depth. Beginning with limited, centrally located electrode positions could risk missing effective and potentially better montages, especially those involving lateral or anterior placements that might enable deeper and more focused modulation within the spinal cord.

In this study, the interferential electric field from a random montage was obtained by a linear combination of two randomly selected electrode pairs, as estimated using Equation (3) ($E_{AM}$). A total of 10,000 montages were randomly selected from a grid of 162 potential positions that surround the thorax in section 3.1. The same number of montages was also selected randomly selected from a reduced grid (44 positions) in the relevant electrode map (see sections 2.5 and 3.2). This was a minimum number of combinations to search for the optimal montage [41] from the total number of combinations. The computation was approximately 2 hours to estimate the $\vec{E}_{AM}$ for all randomly selected montages.

Two metrics were calculated for optimization. The first metric quantifies the neuromodulation effect as the maximum amplitude of the interferential electric field $\vec{E}_{AM}$ in the spinal cord, $\vec{E}_{AM(SC)} = max|\vec{E}_{AM}(SC)|$. The second metric is the focality ratio between the $\vec{E}_{AM}$ in surrounding tissue and that in the spinal cord. Higher focality indicates more selective stimulation of the spinal cord and reduced unintended activation in surrounding tissues, such as skin (perceptual sensations) and muscle (unwanted contractions). It is calculated for the skin as $\vec{E}_{AM(SC/Skin)} = max|\vec{E}_{AM}(SC)|/max|\vec{E}_{AM}(Skin)|$ and for the muscle as: $\vec{E}_{AM(SC/Muscle)} = max|\vec{E}_{AM}(SC)|/max|\vec{E}_{AM}(Skin)|$. Plotting these two metrics for each montage forms a Pareto front, obtained by simultaneously maximizing $\vec{E}_{AM}(SC)/\vec{E}_{AM}(Skin)$ and $\vec{E}_{AM}(SC)$, as shown in Fig. 2B.

2.5. Relevant electrode map

We propose reducing the total number of potential electrode positions by introducing the "Relevant Electrode Map", which shows the most influential positions. Figure 3A presents a 2D map displaying all electrode positions. Figure 3B shows the process used to identify the most relevant electrode positions. Each montage was assigned a score. A score of one point corresponds to montages on the Pareto front (green dot) derived from 10,000 random montages based on the 162-grid. A score of two points was assigned to the best-performing montage among them (red dot). These scores are accumulated over six repetitions and then normalized to generate the final Relevant Electrode Map.

3. Results

3.1. Optimization evaluation

Figure 4 presents the locally optimized tISCS montage using 3D or 2D Pareto approaches. The 2D pareto fronts are obtained by evaluating trade-offs between the target spinal cord intensity $\vec{E}_{AM(SC)}$ and one of the superficial focality metrics, either $\vec{E}_{AM(SC/Skin)}$ or $\vec{E}_{AM(SC/Muscle)}$. The 3D Pareto simultaneously considered all three metrics. Results indicate that superficial electric fields are minimized more effectively using either the 3D Pareto approach or the 2D approach involving skin.

Table 2 compares optimized tISCS with 2-tSCS, which share the same electrode configuration as tISCS. The tISCS approach significantly reduces superficial $E_{AM}$, with reductions of 20.5-fold in the skin and 2.8-fold in the muscle compared to 2-tSCS. While the spinal cord intensity under tISCS decreases by a factor of 1.2 times, the ratios spinal cord to skin and spinal cord to muscle increase by 17.0 and 2.4 times, respectively, compared to 2-tSCS. Table 2 also includes a comparison between the optimized tISCS and conventional tSCS where the total current is the same (i.e, tISCS halves the injection current per electrode compared to tSCS). tISCS reduces peak $E_{AM}$ in the skin and muscle by factors of 23.5 and 5.8, respectively. Similarly, although peak $E_{AM}$ spinal cord decreases by a factor of 2 under tISCS, the spinal cord-to-skin and spinal cord-to-muscle ratios remain higher by 11.3 and 2.9 times, respectively, compared to tSCS.

To assess the robustness of the optimization, we evaluated the consistency of the locally optimized montage across six trials using different subsets of 10,000 randomly selected montages. Table 3 summarizes the findings, demonstrating that the optimized montage consistently reduces superficial electric field intensities relative to conventional tSCS. Across trials, the relative standard deviation of the interferential electric fields ranged from 14% to 27% within tissues, and from 6% to 16% for the focality ratios, confirming the stability of the optimization outcomes.

3.2. Optimization on relevant electrode positions

Considering the variability among different subsets, we focus on identifying the most influential electrode positions to narrow the search space. A Global Relevant Electrode Map was created by integrating data from the six trials, as shown in Fig. 5A, following the procedure outlined in Fig. 3B. The color intensity on the map reflects the frequency of electrode utilization, with more prominent colors indicating positions associated with favorable outcomes. The results show that electrodes in the central dorsal region were predominantly used. As a result, the number of potential electrode locations (search space) was reduced from 162 to 44 considering a threshold over the map of 0.5 covering electrodes on the back.

Figure 5B illustrates the optimization based on the 44-grid. The yellow point represents trial 1 in Section 3.1, which is based on the 162-grid, while the red point indicates the best montage based on the 44-grid, which is closer to the ideal point. Figure 5C displays the peak interferential electric field distribution $E_{AM}$ within the thorax model. As indicated in Table 4, the locally best montage using the 44-grid improved spinal cord intensity by 1.17 times compared to the best 162-grid montage, while enhancing the spinal cord-to-muscle ratio by 1.12 and reducing spinal cord-to-skin ratios by 1.12. These findings suggest that refining electrode positions using the Global Relevant Electrode Map enhances the optimization process for tISCS.

Furthermore, tISCS (44-grid) improves the metrics compared to tSCS and 2-tSCSs. Compared to 2-tSCS, the spinal cord intensity under tISCS decreases by a factor of 1.1 times while the spinal cord-to-skin and spinal cord-to-muscle ratios remain higher by 9.8 and 2.9 times, respectively. Compared to tSCS, the spinal cord intensity under tISCS decreases by a factor of 1.7 times while the spinal cord-to-skin and spinal cord-to-muscle ratios remain higher by 13.0 and 3.7 times, respectively.

## 4. Discussion

Transcutaneous spinal cord stimulation (tSCS) has been investigated for its applications in pain management, motor function rehabilitation, and the treatment of chronic and neuropathic pain. However, tSCS faces challenges in accurately targeting specific nerve regions due to its low focality and significant current attenuation when penetrating deeper areas of the spinal cord. Consequently, higher electrical currents are often needed, which can activate nociceptors and induce pain, while also potentially stimulating nearby organs, including muscles. This study aims to explore the feasibility of using temporal interference for spinal cord neurostimulation through computational analysis, focusing on overcoming these challenges and providing a more precise and effective approach.

The results of our study demonstrate that tISCS reduces off-target skin intensities by at least a factor of 20 compared to tSCS. This reduction is crucial as it minimizes discomfort. Moreover, the spinal cord-to-skin electric field ratio in tISCS is at least 10 times higher than in either 2-tSCS or tSCS. This ratio enables a fivefold increase in spinal cord stimulation intensity without increasing off-target stimulation, even when accounting for the approximately twofold reduction in spinal cord field strength inherent to tISCS. Importantly, with a total injection current of 2.5 mA, tISCS still achieves the minimum electric field threshold for neuromodulation (0.35 V/m) [42]. The optimization was based on a Pareto front representing the trade-off between maximizing the electric field in the spinal cord and increasing the spinal cord-to-off-target tissue ratio. Notably, skin tissue served as a more effective off-target constraint, aligning with the lower activation thresholds of non-nociceptive fibers compared to muscle fibers.

Our approach also addresses the constraint on computation time. Using the lead-field matrix each montage is solved in 80 seconds, which is faster than previously reported times (e.g., 2.5 h per montage, 26 million DOF, COMSOL Multiphysics [19]). Moreover, using a reduced electrode position space (44-grid) enabled a more efficient exploration of potential electrode placements without compromising computational time. It was demonstrated that optimization at more relevant positions improved the trade-off between intensity in the spinal cord and focality.

The current study has several limitations. Major potential inaccuracies in electric field FEM simulations stem from anatomical approximations and assumptions about tissue conductivity. Firstly, our model acts as a proxy for an anatomical model representing the lower thorax, given that simplified models capture the complexities of more detailed models [43]. Future research should consider subject-specific modelling (e.g., based on MRI or other high-resolution images) to investigate inter-individual differences in the effects of tISCS in clinical research applications. In such a case, the leadfield approach should be applied on a subject-specific basis, and integrating it with other optimization methods could further accelerate the process [44]. Secondly, while the selection of conductivity values may affect absolute peak magnitudes, it does not significantly influence the relative spatial distribution of the electric field and is therefore unlikely to hinder the optimization outcome [45]. Finally, we relied on the quasi-static assumptions, which include neglecting dispersion phenomena, and treating the medium as purely resistive without considering capacitive effects. Relaxing the quasi-static assumptions can result in differences of up to 20% in model fiber responses to kilohertz stimuli [49]. Therefore, more precise estimation of the temporal dynamics of the electric field may be determinant for optimization strategies focused on combined assessment of electrode placement and stimulation waveform [46] [47].

## 5. Conclusion

This study presents a computational analysis aimed at optimizing transcutaneous interference spinal cord stimulation (tISCS). Results show that tISCS significantly reduces off-target stimulation in superficial tissues and enhances spinal cord-to-skin connectivity compared to conventional tSCS. A leadfield matrix alleviated computational challenges and shortened simulation time, enabling the evaluation of an extensive electrode position search that could be improved by adopting a refined montage space based on more influential positions. This study advances tISCS as a promising non-invasive alternative to traditional spinal cord stimulation, allowing for stronger stimulation without compromising patient comfort.

**Acknowledgement**

This study was supported by a JSPS Grant-in-Aid for Scientific Research, JSPS KAKENHI Grant-25K15887, and ANID Fondecyt Iniciación 11190822. We thank Scott Lempka for providing us with the spinal model on which we based the development of this transcutaneous model.

**Conflict of Interest**

None

**Tables and Figures**

**Table 1.** Tissue electrical conductivity in the human thorax model

| Tissue | Conductivity (S/m) | Reference |
|---|---|---|
| Skin | 0.148 | [27] |
| Fat | 0.077 | [27] |
| Muscle | 0.355 | [48] |
| General thorax (water) | 0.250 | [49] |
| Bone / Vertebrae | 0.006 | [27] |
| Intervertebral disks | 0.650 | [29] |
| Extradural tissue | 0.250 | [29] |
| Dura matter | 0.461 | [30] |
| CSF | 1.880 | [27] |
| Spinal-WM (transverse) | 0.083 | [31] |
| Spinal-WM (longitudinal) | 0.600 | [31] |
| Spinal-GM | 0.230 | [31] |

**Table 2** Comparison between stimulation methods and montage optimization

| Metrics | tSCS | 2-tSCS | tISCS (2D) SC/Skin | tISCS (2D) SC/Muscle | tISCS (3D) |
|---|---|---|---|---|---|
| (a) Electric Field [V/m] | | | | | |
| Spinal cord | 4.24 | 2.54 | 2.15 | 2.67 | 2.15 |
| Skin | 148.30 | 129.40 | 6.32 | 104.70 | 6.32 |
| Muscle | 22.07 | 10.88 | 3.82 | 4.77 | 3.82 |
| (b) Ratio | | | | | |
| Spinal cord / Skin | 0.03 | 0.02 | 0.34 | 0.02 | 0.34 |
| Spinal cord / Muscle | 0.19 | 0.23 | 0.56 | 0.56 | 0.56 |

**Table 3** Consistency of locally optimized montages

| Metrics | tSCS | tISCS 1st | tISCS 2nd | tISCS 3rd | tISCS 4th | tISCS 5th | tISCS 6th | Relative SD |
|---|---|---|---|---|---|---|---|---|
| (a) Electric Field [V/m] | | | | | | | | |
| Spinal cord | 4.23 | 2.15 | 1.39 | 1.78 | 1.47 | 1.71 | 1.39 | 16% |
| Skin | 148.30 | 6.32 | 3.86 | 6.83 | 3.33 | 4.65 | 4.0 | 27% |
| Muscle | 22.07 | 3.82 | 2.66 | 2.88 | 2.73 | 2.91 | 2.62 | 14% |
| (b) Ratio | | | | | | | | |
| Spinal cord / Skin | 0.03 | 0.34 | 0.36 | 0.26 | 0.44 | 0.37 | 0.41 | 16% |
| Spinal cord / Muscle | 0.19 | 0.56 | 0.53 | 0.62 | 0.54 | 0.59 | 0.53 | 6% |

**Table 4** The comparison of E-field magnitude between tSCS and tISCS

| Metrics | tSCS | 2-tSCS | tISCS (44-grid) | tISCS (162-grid) |
|---|---|---|---|---|
| Electric field in spinal cord (V/m) | 4.24 | 2.67 | 2.51 | 1.39-2.15 |
| Spinal cord / Skin | 0.03 | 0.04 | 0.39 | 0.26-0.44 |
| Spinal cord / Muscle | 0.19 | 0.24 | 0.70 | 0.53-0.62 |

**Fig. 1. Human and electrode model.** (A) The whole thorax model and each tissue. (B) The model of the spinal cord. (C) The model of electrode. (D) Transcutaneous interference spinal cord stimulation, or tISCS.

**Fig. 2. Optimisation procedure**. (A) A total of 162 electrodes are used to build the leadfield matrix. (B) Amplitude-modulated envelope ($E_{AM}$) are obtained from a total of 10,000 random electrode combinations using the leadfield matrix. A Pareto front is formed from the trade-off between the maximum intensity of $E_{AM}$ in the spinal cord and the ratio of $E_{AM}$ between the spinal cord and skin (or muscle).

**Fig. 3. Influential electrode positions in the grid.** (A) Representation of electrode positions in a plane. (B) Outline to obtain the Relevant Electrode Map.

**Fig. 4. tISCS optimal montage (162-grid).** Local best montage from a total of 10,000 random selected montages based on Pareto fronts (2D and 3D). As reference a common employed tSCS is shown.

**Fig. 5. tISCS optimal montage with reduced grid space (44-grid).** (A) Most dominant electrode positions. (B) Local best montage from a total of 10,000 random selected montages based on Pareto fronts (3D). The same results of the 3D pareto front is also shown from two 2D perspectives.

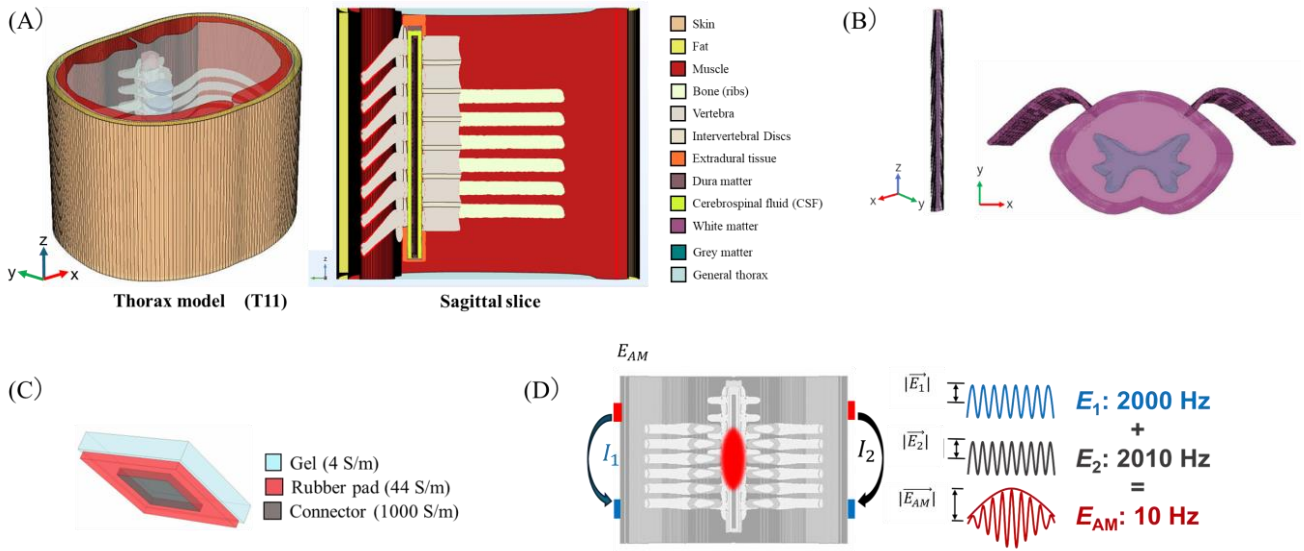

**Figure 1**

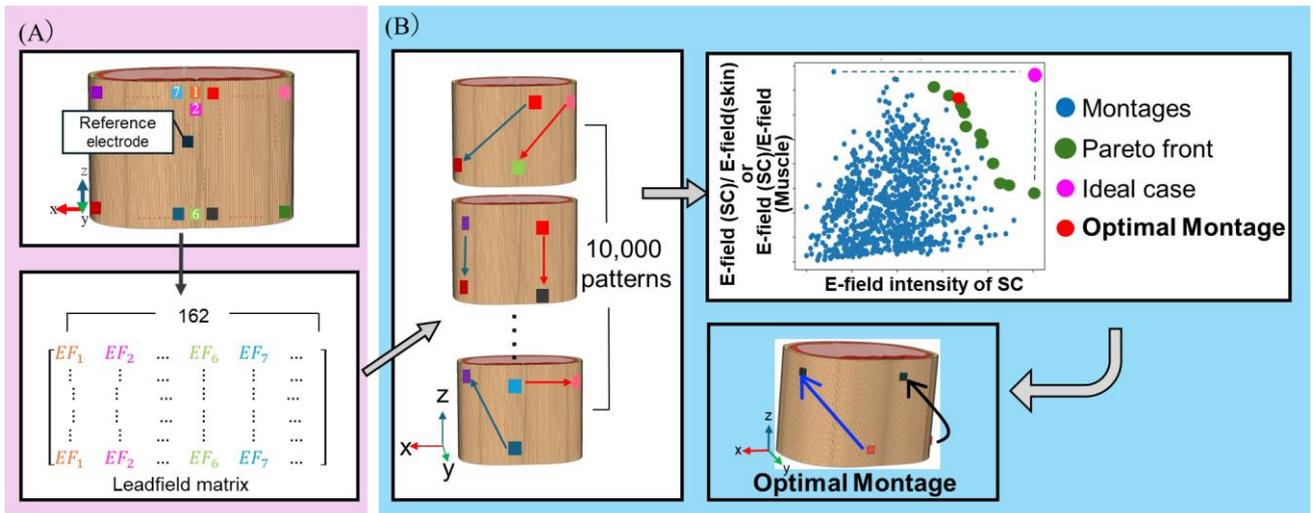

**Figure 2**

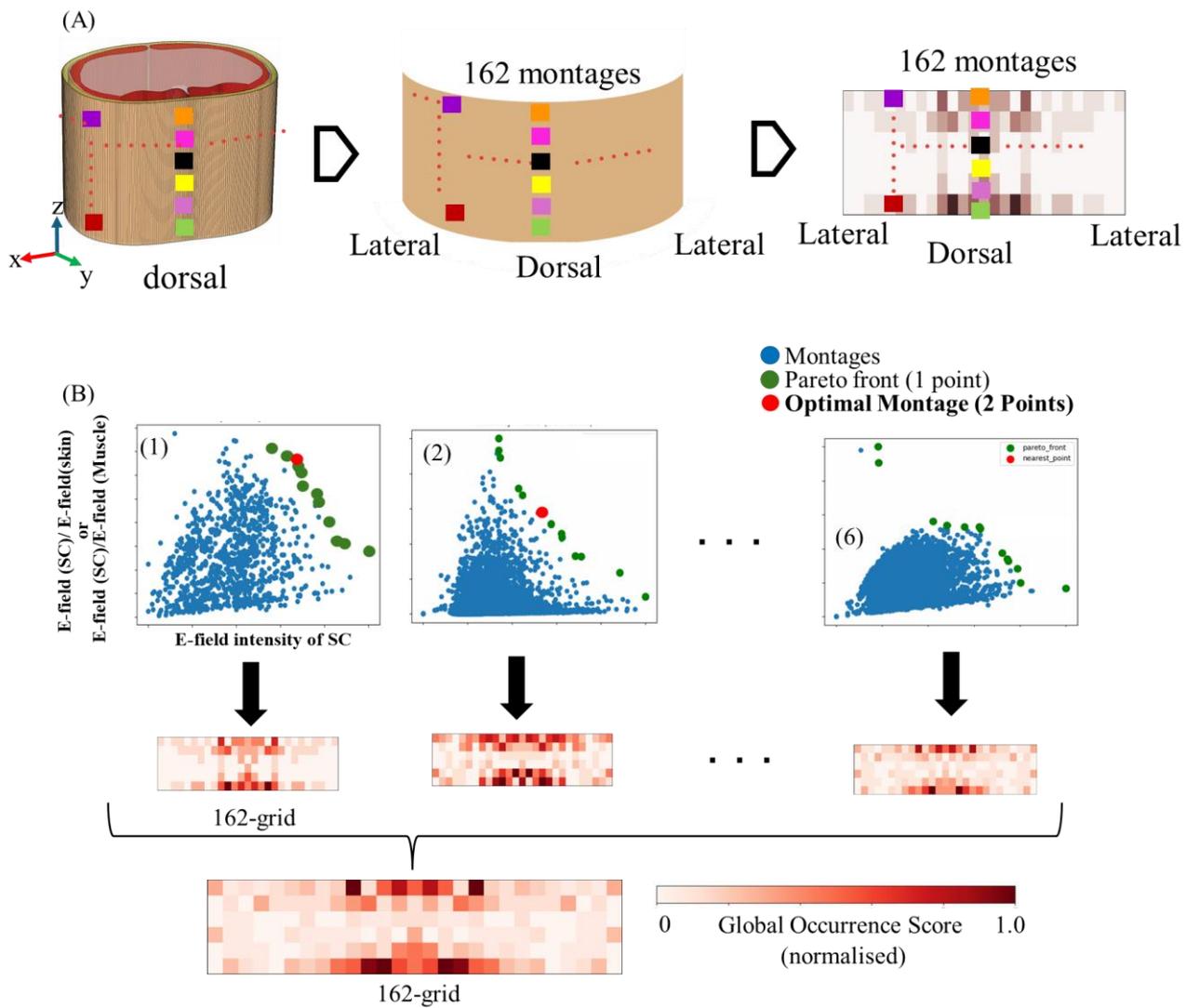

**Figure 3**

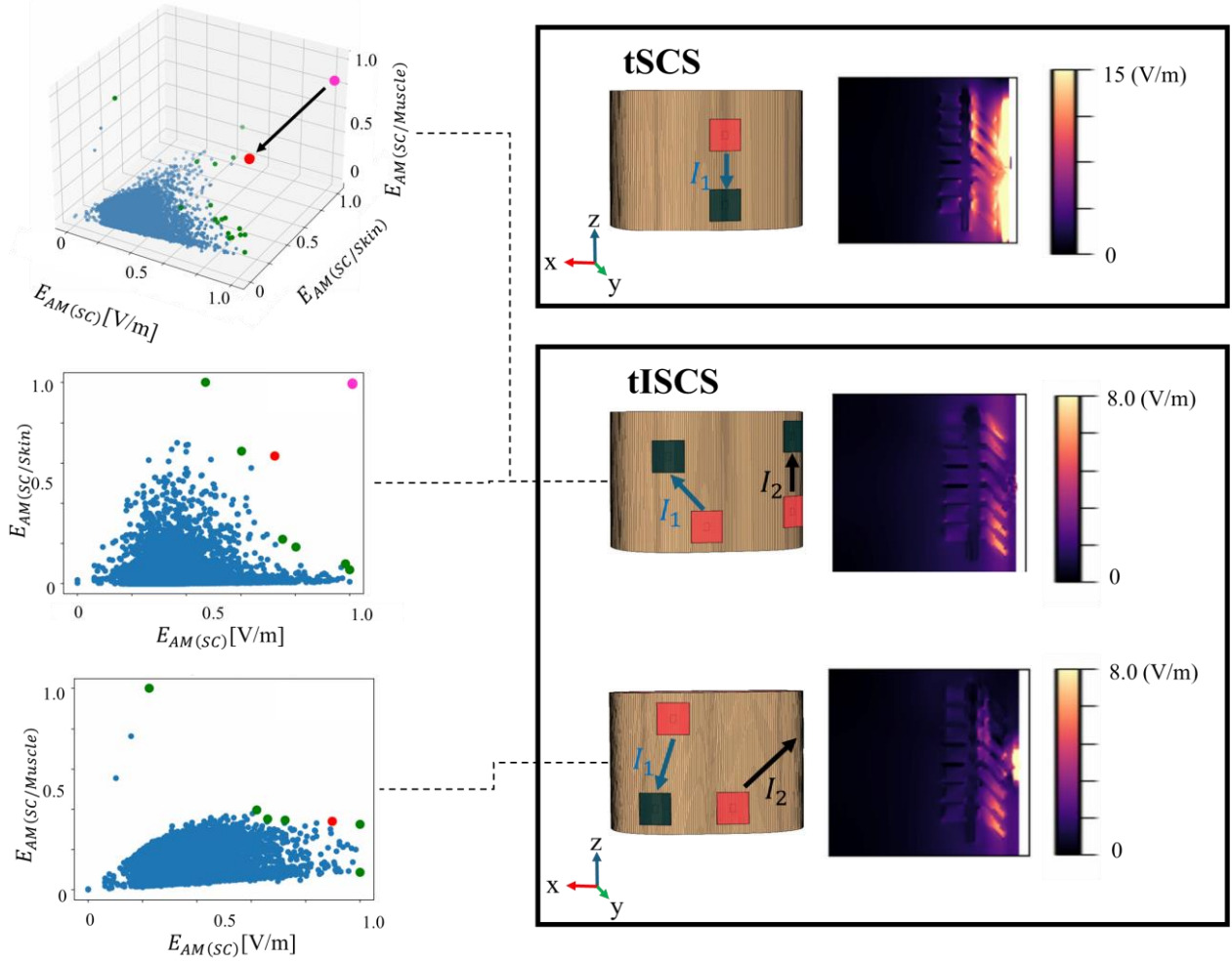

**Figure 4**

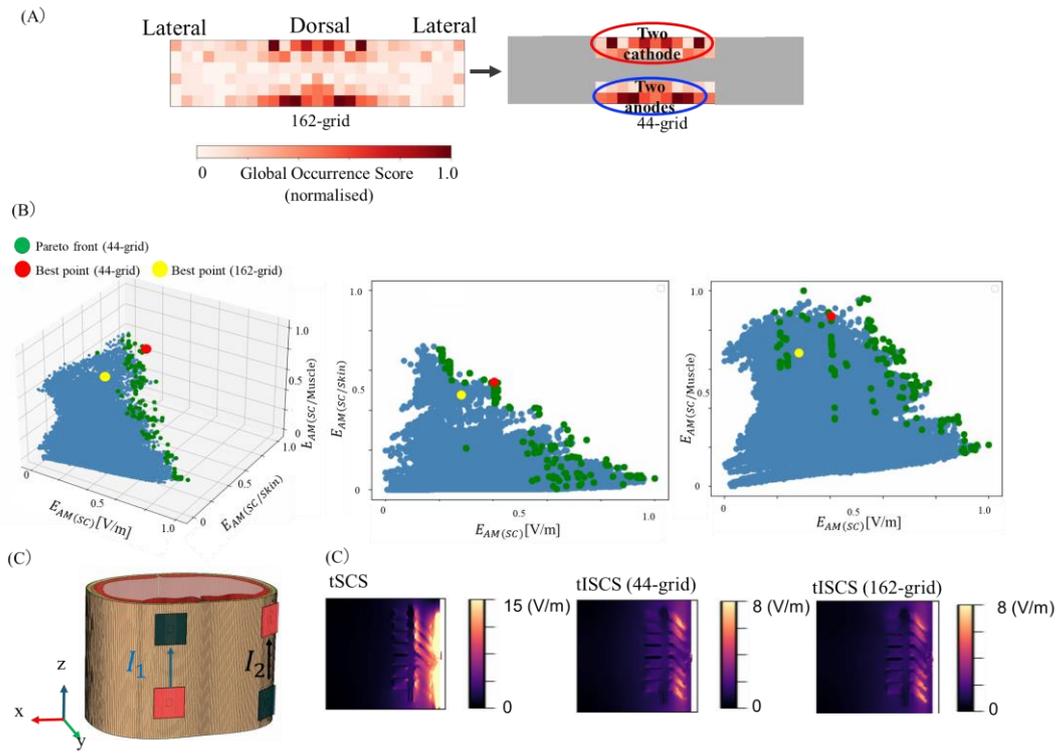

**Figure 5**